\documentclass{aastex}
\usepackage{emulateapj5}

\begin{document}

\def\gsim{\lower0.5ex\hbox{$\; \buildrel > \over \sim \;$}}
\def\lsim{\lower0.5ex\hbox{$\; \buildrel < \over \sim \;$}}
\def\rs{\mbox{$R_{\rm s}$}}
\def\rms{\mbox{$R_{\rm ms}$}}
\def\rh{\mbox{$R_{\rm h}$}}
\def\ro{\mbox{$R_0$}}
\def\ak{\mbox{$a_{\rm k}$}}
\def\nuu{\mbox{$\nu_{\rm u}$}}
\def\nuk{\mbox{$\nu_{\rm k}$}}
\def\num{\mbox{$\nu_{\rm m}$}}
\def\nub{\mbox{$\nu_{\rm b}$}}
\def\dnu{\mbox{$\Delta\nu$}}
\def\ms{\mbox{$M_{\odot}$}}
\def\etal{\mbox{\it et al.}}
\newcommand{\gcc}{g~cm$^{-3}\ $}
\newcommand{\sfun}[2]{$#1(#2)\ $}
\newcommand{\rhonot}{$\rho_{\circ}\ $}
\newcommand{\msun}{$M_{\odot}\ $}
\newcommand{\greq}{$\stackrel{>}{ _{\sim}}$}
\newcommand{\lteq}{$\stackrel{<}{ _{\sim}}$}

\def\be{\begin{equation}}
\def\ee{\end{equation}}

\title{Rapidly rotating strange stars for a new equation of state 
of strange quark matter} 
\author{ Ignazio Bombaci\altaffilmark{1},
         Arun V. Thampan\altaffilmark{2}, and  
         Bhaskar Datta\altaffilmark{3,4} }
\altaffiltext{1}{ Dipartimento di Fisica, Universit\'a di Pisa, 
and INFN Sezione Pisa, via Buonarroti 2, I--56127, Italy;
BOMBACI@pisa.infn.it}
\altaffiltext{2}{Inter--University Centre for Astronomy and Astrophysics 
                 (IUCAA), Pune 411 007, India; arun@iucaa.ernet.in } 
\altaffiltext{3}{Indian Institute of Astrophysics, Bangalore 560034, India}
\altaffiltext{4}{Raman Research Institute, Bangalore 560080, India  } 
\begin{abstract}
For a new equation of state of strange quark matter, we construct equilibrium 
sequences of rapidly rotating strange stars in general relativity.  
The sequences are the normal and supramassive evolutionary sequences 
of constant rest mass.  We also calculate equilibrium sequences 
for a constant value of $\Omega$ 
corresponding to the most rapidly rotating pulsar PSR~1937~$+$21.  
In addition to this, we calculate the radius of the marginally 
stable orbit and its dependence on $\Omega$, relevant for modeling of 
kilo--Hertz quasi--periodic oscillations in X--ray binaries.
\end{abstract}   
\keywords{stars: neutron - equation of state - general relativity}

\section{Introduction}
One of the most fascinating aspects of modern astrophysics,  
is the possible existence of a new family of collapsed stars   
consisting completely of a deconfined mixture of {\it up} ({\it u}),  
{\it down} ({\it d}), and {\it strange} ({\it s}) quarks, together with 
an appropriate number of electrons to guarantee electrical neutrality. 
Such compact stars have been referred to in the literature, as strange 
stars (SS), and their 
constituent matter as strange quark matter (SQM). 
The possible existence of SS is a direct consequence of the 
the so called {\it strange matter hypothesis}, formulated by Witten~(1984) 
(see also Bodmer 1971, Terazawa 1979).         
According to this hypothesis, SQM, in equilibrium with 
respect to the weak interactions, could be the absolute ground state of 
strongly interacting matter rather than $^{56}$Fe.  
Ever since the formulation of this hypothesis, there have been several
reports in the literature, on the structure of non--rotating SS  (e.g. 
Haensel, Zdunik \& Schaefer 1986; Alcock, Farhi \& Olinto 1986).   
However, these have remained largely theoretical and speculative in nature. 
Observational data from RXTE of certain low mass X--ray binaries (LMXBs) 
have recently given rise to a resurgence in the astrophysical interest 
associated with SS.  It has been suggested that at least a few LMXBs could be 
harboring SS. For example, recent studies have shown that the compact objects  
associated with the X--ray burster 4U~1820~$-$30 (Bombaci 1997), the bursting 
X--ray pulsar GRO~J1744~$-$28 (Cheng \etal\ 1998) and the X--ray pulsar 
Her~X$-$1 (Dey \etal\ 1998) are likely SS candidates.    
The most promising and convincing SS candidates are the compact 
objects in the bursting millisecond X--ray pulsar SAX~J1808.4~$-$3658 
(Li \etal\ 1999a), and in the atoll source 4U~1728~$-$34 (Li \etal\ 1999b). 

The compact nature of these sources make general relativity important in
describing these systems. Furthermore, their existence in binary systems 
imply that these may possess rapid rotation rates (Bhattacharya \& van den
Heuvel 1991 and references therein).  
Particularly, the two SS candidates in SAX~J1808.4~$-$3658 and 
4U~1728~$-$34 are millisecond pulsars having spin periods $P=2.49$~ms and 
$P=2.75$~ms respectively.  These two properties make the incorporation 
of general relativistic effects of rotation imperative for satisfactory 
treatment of the problem.   

Most of the calculations on the rotational properties of SS, reported
so far, have relied on the slow 
rotation approximation (Colpi \& Miller 1992, Glendenning \& Weber 1992). 
This approximation loses its validity as the star's spin frequency 
approaches the mass shedding limit. 
Rapidly rotating SS sequences have been recently  
computed by Gourgoulhon \etal\ (1999), Stergioulas \etal\ (1999), and 
Zdunik \etal\ (2000). 
However, all the calculations mentioned above make use of a very schematic 
model (Freedman \& McLerran 1978, Farhi \& Jaffe, 1984), related to the MIT 
bag model (Chodos \etal\ 1974) for hadrons, for the equation of state (EOS) 
of SQM. 
Within the MIT bag model EOS, the SS radii calculated are seen 
to be incompatible with the mass--radius ($M$--$R$) relation (Li \etal\ 1999a)
for SAX~J1808.4~$-$3658, and only marginally compatible (see Fig. 2)
with that for 4U~1728~$-$34 (Li \etal\ 1999b).
 
In this letter, we present calculations of equilibrium sequences 
of rapidly rotating 
SS in general relativity using a new model for the EOS 
of SQM derived by Dey \etal\ (1998). This model is based on a 
``dynamical'' density--dependent approach to confinement. In contrast,
in the simple EOS based on the MIT bag model, the medium effects on the
quark degrees of freedom and on the quark--quark interaction are not 
considered.  For illustrative purposes, we compare our results, obtained 
using this new EOS, with those for 
the MIT bag model  EOS.

We use the methodology described in detail in Datta, Thampan \& Bombaci (1998)
to calculate the structure of rapidly rotating SS.  
For completeness, we briefly describe the method here.  For a general 
axisymmetric and stationary space--time, assuming a perfect fluid 
configuration, the Einstein field equations reduce to ordinary integrals
(using Green's function approach).  These integrals may be self consistently
(numerically and iteratively) solved to yield the value of metric coefficients 
in all space.  Using these metric coefficients, one may then compute the 
structure parameters, angular momentum and moment of inertia corresponding to 
initially assumed central density and polar to equatorial radius ratio.  
These may then be used (as described in Thampan \& Datta 1998) to 
calculate parameters connected with stable circular orbits (like the innermost 
stable orbit and the Keplerian angular velocities) around the configuration 
in question.

The sequences that we calculate are: constant rest mass sequences, constant 
angular velocity sequences, constant central density sequences and constant 
angular momentum sequences.   
We also calculate the radius $r_{\rm orb}$ of the marginally stable orbit 
and its dependence on the spin rate of the SS, which will be
relevant for modeling X--ray burst sources involving SS.

\section{The equation of state for SQM}
As mentioned earlier, the schematic EOS for SQM 
based on the MIT bag model, has become the ``standard'' EOS model for SS 
studies. However, this EOS model becomes progressively less trustworthy  
as one goes from very high density region (asymptotic freedom regime) to that 
of low densities, where confinement (hadron formation) takes place.  
Recently, Dey \etal\ (1998) derived a new EOS for SQM  using a 
``dynamical'' density--dependent approach to confinement.  
This EOS 
has asymptotic freedom built in, shows 
confinement at zero baryon density, deconfinement at high density. 
In this model, the quark interaction is described by a colour--Debye--screened  
inter--quark vector potential originating from gluon exchange, 
and by a density--dependent scalar potential which restores chiral 
symmetry at high density (in the limit of massless quarks).    
This density--dependent scalar potential arises from the density 
dependence of the in--medium effective quark masses $M_{\rm q}$, which 
are taken to depend on the baryon number density 
$n_{\rm B}$ according to (see Dey \etal\ 1998)
\be 
M_{\rm q} = m_{\rm q} + 310 \cdot sech\bigg(\nu {{n_{\rm B}}\over{n_0}}\bigg)
                                   \qquad  \qquad  ({\rm MeV}),    
\ee
where $n_0 = 0.16$~fm$^{-3}$ is the normal nuclear matter density,  
${\rm q (= u,d,s)}$ is the flavor index, and $\nu$ is a parameter.   
The effective quark mass $M_{\rm q}(n_{\rm B})$ goes from its constituent 
mass at 
$n_{\rm B}=0$, to its current mass $m_{\rm q}$,  as 
$n_{\rm B} \rightarrow \infty$.
Here, we consider two different parameterizations of this EOS, 
which correspond to a different choice for the parameter $\nu$. 
The  equation of state SS1 (SS2) corresponds to $\nu = 0.333$ ($\nu = 0.286$). 
These two models for the EOS give absolutely stable SQM according to the 
strange matter hypothesis. 

In order to compare our results with those of
previous studies, we also use the MIT bag model EOS, for massless
non--interacting quarks and $B=90$~MeV/fm$^3$ (hereafter the $B90\_0$
EOS).

\section{Results and discussion}
The equilibrium sequences of rotating SS depend on two parameters: 
the central density ($\rho_{\rm c}$) and the rotation 
rate ($\Omega$).  For purpose of illustration, we choose three limits
in this parameter space. These are: (i) the static or non--rotating limit, 
(ii) the limit at which instability to quasi--radial mode sets in 
and (iii) the centrifugal mass shed limit. The last limit corresponds to the 
maximum $\Omega$ for which centrifugal forces are able to balance the inward 
gravitational force.

The result of our calculations for EOS SS1 is displayed in Fig. 1.
In Fig. 1 (a) we show the functional dependence of the gravitational 
mass ($M$) with  $\rho_c$.  In these set of figures, the bold solid 
curve represents the non--rotating or static limit, and the bold dashed 
curve the centrifugal mass shed limit.  
The thin solid curves are the constant rest mass ($M_0$) 
evolutionary sequences. 
The evolutionary sequences above the maximum stable non--rotating mass
configuration are the supramassive evolutionary sequences, and those that 
lie below this limit are the normal evolutionary sequences.  
The maximum mass sequence for this EOS  corresponds to $M_0 = 2.2$~\msun.  
The thin dot--dashed line (slanted towards left) represents 
instability to quasi--radial perturbations.  
In the central panel of the same figure, we give a plot of $M$ as a 
function of the equatorial radius $R$.    

In panel (c) of Fig. 1,  we display the plot of $\Omega$ as a function of 
the specific angular momentum $\tilde{j}=cJ/GM_{0}^{2}$  
(where $J$ is the angular momentum  of the configuration).  
Unlike for neutron stars (e.g. Cook, Shapiro \& Teukolsky 1994; 
Datta, Thampan \& Bombaci 1998), the $\Omega$--$\tilde{j}$ curve
does not show a turn--over to lower $\tilde{j}$ values for SS.  
This is due to the effect of the long--range (non--perturbative) interaction 
in QCD (Quantum Chromo--Dynamics), which is responsible for quark confinement 
in hadrons, and makes low mass SS self--bound objects.  
$\Omega$ for the mass shed limit appears to asymptotically 
tend to a non--zero value for rapidly rotating low mass stars.  
A further ramification of this result is that the ratio of the rotational 
energy to the total gravitational energy ($T/W$) becomes greater than 0.21 
(as also reported by Gourgoulhon \etal\ 1999) thus probably making the
configurations susceptible to triaxial instabilities.

In Table 1 we display the values of the structure parameters for the
maximum mass non--rotating SS models.  
The larger value of the maximum mass for the SS1 model, with respect to 
the SS2 model, can be traced back to role of the parameter $\nu$ 
in eq. (1) for the effective quark mass $M_q$.  
In fact, a larger value of $\nu$ (SS1 model) gives a faster decrease 
of $M_q$ with density, producing a stiffer EOS.   

Table 2 and Table 3 displays the maximum mass rotating and maximum
angular momentum models for the EOS models under consideration.  
While for EOS SS1, the maximum mass rotating model and the maximum
angular momentum models are the same, for EOS SS2, the two models are
slightly different, with the maximum angular momentum model coming
earlier (with respect to $\rho_{\rm c}$) than the maximum mass
rotating configuration.

In Table 4 we list the values of the various parameters for the constant
$\Omega$ sequences for EOS SS1.  
The first entry in this table corresponds to $\rho_{\rm c}$ for which
$r_{\rm orb}=R$.  For higher values of $\rho_{\rm c}$, $r_{\rm orb}>R$;
for large values of $\rho_{\rm c}$, the boundary layer (separation between 
the surface of the SS and its innermost stable orbit) can be substantial 
($\sim 5$ km for the maximum value of the listed $\rho_{\rm c}$) . 

In Fig. 2, we plot our theoretically calculated $M$--$R$ curves 
(dashed curves) for EOS SS1 and B90\_0, for a rotational frequency of  
364~Hz  corresponding to the inferred rotational frequency of the compact 
star in the source 4U~1728~$-$34 (M\'{e}ndez \& Van der Klis 1999).  
In the same figure, we plot the radius $R_0$ of the inner edge of the 
accretion disk (dotted curve) for 4U~1728~$-$34 as deduced by Titarchuk 
\& Osherovich (1999) (see also Li et al. 1999b) from a fit of kHz QPO data 
in this source.  Since $R_0$ must be larger than both  $R$ and $r_{\rm orb}$, 
it is possible to deduce (Li et al. 1999b) an upper bound for the radius and 
mass of the central  accretor. Using these constraints on M and R, 
Li et al. (1999b) concluded that 4U~1728~$-$34 is possibly a strange star 
rather than a neutron star. 
However, in their calculation, Li et al. (1999b) used an approximate 
EOS--independent expression to account for the effects of rotation on the 
moment of inertia of the star and hence on $r_{\rm orb}$ (see Li et al 1999b 
for further details).  In contrast, in this letter we make an ``exact'' 
calculation (as a result of our accurate calculation of angular momentum and 
hence the moment of inertia) of $r_{\rm orb}$ (triple--dot dashed curves) 
for the two EOS SS1 and B90\_0. 
Notice that the upper bound for the mass of 4U~1728~$-$34 (the intersection 
point between the curves $R_0$ and $r_{\rm orb}$) is about 1.15~$M_\odot$, 
in agreement with the results of Li et al. (1999b). 
The two EOS models considered in Fig. 2 are both consistent with 
4U~1728~$-$34.  
Also, having in mind the scaling with $B^{-1/2}$  of the M--R relation 
of SS within the bag model EOS (Witten, 1984) and the constraints on the 
allowed values of the constant $B$ to fulfil the strange matter hypothesis 
(Farhi \& Jaffe, 1984),  we see that the simple EOS based on the MIT bag 
model is only marginally compatible with the M--R relation for 4U~1728~$-$34. 

To summarize, in this letter, we present calculations of equilibrium
sequences of rapidly rotating strange stars in general relativity, for
a set of new equations of state.  We compare the results so obtained
with those for the generally used MIT bag model EOS.  In addition to
this, we illustrate the results of our computations for a specific
rotation rate inferred for the central accretor in the source 
4U~1728~$-$34. We find that if the compact star in this source
were indeed to be a strange star, then its mass and radius would
be bounded above by a value of $M=1.15$~\msun and $R=9.3$~km.
It is expected that future observations may shed more light on this issue.




\input psbox.tex

\begin{table*}
\begin{center}
\caption{ Structure parameters for the non--rotating maximum mass
configurations. Listed, are the central density ($\rho_c$) in units of
g~cm$^{-3}$, the gravitational mass ($M$) in solar units, the equatorial 
radius ($R$) in km,  the baryonic mass ($M_0$) in solar units, the radius 
of the marginally stable orbit ($r_{\rm orb}$) in km and the moment of 
inertia ($I$) in units of $10^{45}$~g~cm$^2$.}
\vspace{.4 cm}
\label{tab1}
\begin{tabular}{ccccccc}                          \hline\hline
\multicolumn{1}{c}{EOS} & \multicolumn{1}{c}{$\rho_{\rm c}$} & 
\multicolumn{1}{c}{$M$} & \multicolumn{1}{c}{$R$} & 
\multicolumn{1}{c}{$M_0$} & 
\multicolumn{1}{c}{$r_{\rm orb}$} & \multicolumn{1}{c}{$I$} \\ 
\hline
\\
SS1    &  4.65E+15 &  1.438 &  7.093 & 1.880 &  12.729   &  0.736 \\
SS2    &  5.60E+15 &  1.324 &  6.533 & 1.658 &  11.716   &  0.578 \\
B90\_0 &  3.09E+15 &  1.603 &  8.745 & 1.937 &  14.020   &  1.099 \\
\\
\hline
\end{tabular}
\end{center}
\end{table*}

\vfil\eject
\begin{table*}
\begin{center}
\caption{Structure parameters for the maximally rotating 
($\Omega=\Omega_{\rm ms}$) maximum mass configuration. In addition to
the quantities listed in the previous table, we display, the rotation
rate ($\Omega$) in $10^4$~rad~s$^{-1}$, the ratio of the rotational to the
total gravitational energy ($T/W$) and the specific angular momentum 
($\tilde{j}=cJ/GM_{0}^{2}$).}
\vspace{.4 cm}
\label{tab2}
  \begin{tabular}{cccccccccc}                \hline\hline
\multicolumn{1}{c}{EOS} & \multicolumn{1}{c}{$\rho_{\rm c}$} & 
\multicolumn{1}{c}{$\Omega$} & \multicolumn{1}{c}{$I$} & 
\multicolumn{1}{c}{$M$} & \multicolumn{1}{c}{$T/W$} & 
\multicolumn{1}{c}{$R$} & 
\multicolumn{1}{c}{$\tilde{j}$} & 
\multicolumn{1}{c}{$r_{orb}$} & 
\multicolumn{1}{c}{$M_0$} \\
\hline
\\
SS1 & 3.10E+15  &  1.613 &  2.072 &  2.077 &  0.219  &  10.404  & 
0.524 & 11.656  &  
2.694 \\
SS2 & 3.60E+15  &  1.738 &  1.613 &  1.904 &  0.218  &   9.612  & 
0.570 & 10.758  &  
2.366 \\
B90\_0 & 1.90E+15  &  1.190 &  3.369 &  2.272 &  0.232 &   13.213 &  
0.633 & 14.612  &  
2.683 \\
\\
\hline
\end{tabular}
\end{center}
\end{table*}

\vfil\eject
\begin{table*}
\begin{center}
\caption{Structure parameters for the maximum angular momentum
         configuration.}
\vspace{.4 cm}
\label{tab3}
 \begin{tabular}{cccccccccc}                \hline\hline
\multicolumn{1}{c}{EOS} & \multicolumn{1}{c}{$\rho_{\rm c}$} & 
\multicolumn{1}{c}{$\Omega$} & \multicolumn{1}{c}{$I$} & 
\multicolumn{1}{c}{$M$} & \multicolumn{1}{c}{$T/W$} & 
\multicolumn{1}{c}{$R$} & 
\multicolumn{1}{c}{$\tilde{j}$} & 
\multicolumn{1}{c}{$r_{orb}$} & 
\multicolumn{1}{c}{$M_0$} \\
\hline
\\
SS1 & 3.10E+15  &  1.613 &  2.072 &  2.077 &  0.219  &  10.404  & 
0.524 & 11.656  &  
2.694 \\
SS2 & 3.40E+15 &  1.719 & 1.633 &  1.899 &  0.220  &  9.693  & 
0.575 &  10.837 &  
2.355 \\
B90\_0 & 1.70E+15 &  1.161 & 3.456 &  2.254 &  0.239  & 13.447  & 
0.650 &  14.864 &  
2.650 \\
\hline
\end{tabular}
\end{center}
\end{table*}

\vfil\eject
\begin{table*}
\begin{center}
\caption{Structure parameters for the constant angular velocity 
sequence for EOS SS1. This sequence corresponds to the rotation rate of the 
pulsar PSR~1937~$+$21 (Backer \etal\ 1982), having  
$\Omega=4.03\times10^3$~rad~s$^{-1}$ or period $P=1.556$~ms.}
\vspace{.4 cm}
\label{tab4}
 \begin{tabular}{cccccccc}                \hline\hline
\multicolumn{1}{c}{$\rho_{\rm c}$} & 
\multicolumn{1}{c}{$I$} & 
\multicolumn{1}{c}{$M$} & \multicolumn{1}{c}{$T/W$} & 
\multicolumn{1}{c}{$R$} & 
\multicolumn{1}{c}{$\tilde{j}$} & 
\multicolumn{1}{c}{$r_{orb}$} & 
\multicolumn{1}{c}{$M_0$} \\
\hline
\\
1.70E+15 &  
0.353 &  0.852 &  0.013 &  6.663 &  
0.153 &   6.663 &  
1.027 \\
1.80E+15 &  
0.433 &  0.963 &  0.012 &  6.884 &  
0.143 &   7.664 &  
1.178 \\
1.90E+15 &  
0.502 &  1.052 &  0.019 &  7.036 &  
0.136 &   8.378 &  
1.301 \\
2.40E+15 &  
0.701 &  1.297 &  0.010 &  7.326 &  
0.116 &  10.355 &  
1.660 \\
2.60E+15 &  
0.737 &  1.346 &  0.010 &  7.344 &  
0.112 &  10.741 &  
1.735 \\
4.60E+15 &  
0.769 &  1.458 &  0.008 &  7.139 &  
0.096 &  11.732 &  
1.914 \\
5.65E+15 &  
0.734 &  1.449 &  0.007 &  7.007 &  
0.093 &  11.703 &  
1.899 \\
\\
\hline
\end{tabular}
\end{center}
\end{table*}

\begin{figure*}
\begin{center}
\caption{Structure parameters for rotating strange stars corresponding
to EOS SS1. The bold-solid line represents the non--rotating limit,
the bold-dotted line the mass--shed limit and the almost vertical thin 
dot--dashed line tilted to the left is the instability limit to
quasi--radial mode perturbations.  The thin solid lines (labelled 1, 2
...) represent constant rest mass sequences: 1: 1.59 \msun, 
2: 1.66 \msun, 3: 1.88 \msun, 4: 2.14 \msun 5: 2.41 \msun. }
\hspace{-2.0cm}
{\mbox{\psboxto(19cm;18cm){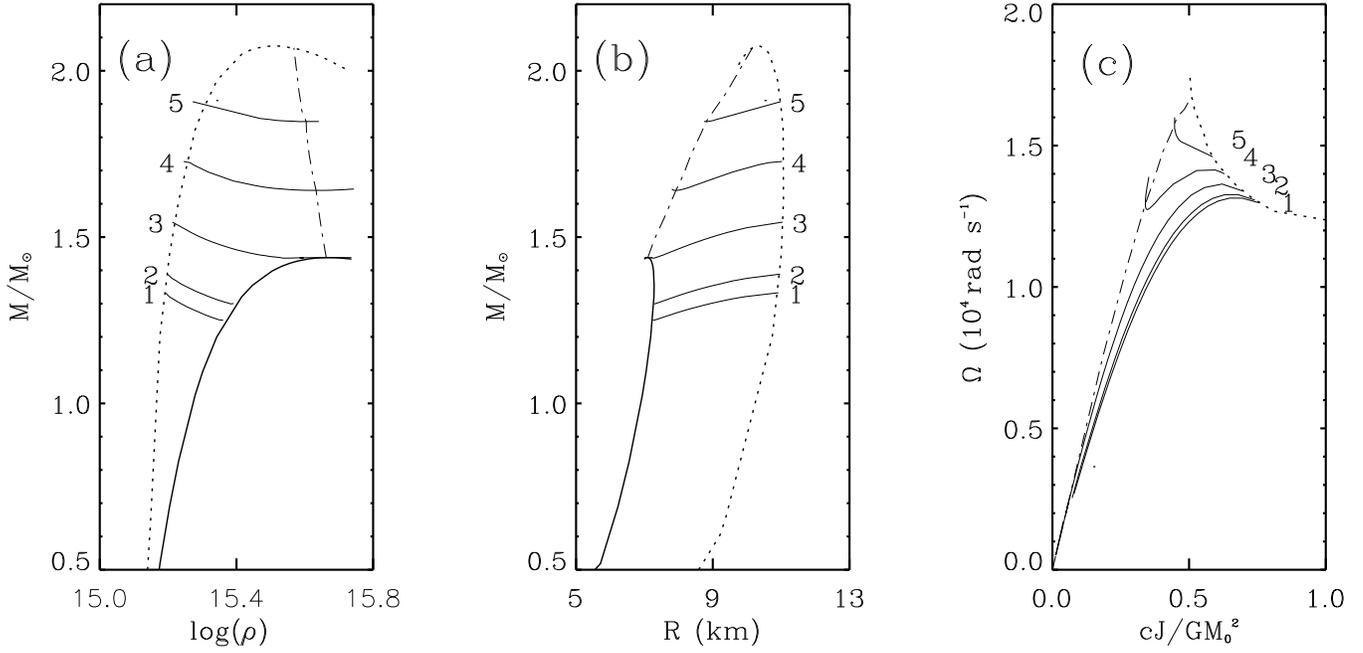}}} 
\end{center}
\end{figure*}

\begin{figure}
\begin{center}
\caption{Mass--radius relations and $r_{\rm orb}$  for a rotation rate of
$\Omega=2.287\times10^3$~rad~s$^{-1}$ (i.e. $364$~Hz) corresponding 
to the inferred rotation rate of the central accretor in 4U~1728~$-$34
(M\'{e}dez and Van der Klis 1999).  Also plotted are the bounds 
on the $M$--$R$ from general considerations, and the fit for these
values obtained by Titarchuk \& Osherovich (1999). See text for
further details.}
\hspace{-2.0cm}
{\mbox{\psboxto(9cm;18cm){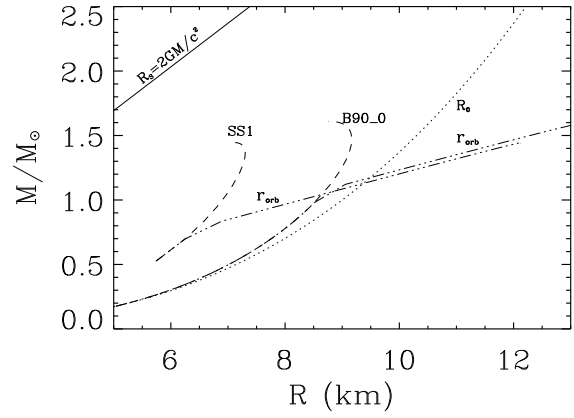}}} 
\end{center}
\end{figure}
\end{document}